# Mendeleev and the Periodic Table of Elements

Subhash Kak

## Abstract

This note presents reasons why Mendeleev chose Sanskrit names (now superseded) for eight elements in the periodic table.

## 1. Introduction

It is an amusing sidelight of history of science that the original names used by Mendeleev for gallium and germanium are eka-aluminum and eka-silicon, where the *eka*, Sanskrit for *one*, has the sense of *beyond*. The prediction for the existence of these elements was made by Mendeleev in a paper in 1869, and it was the identification of these elements in 1875 and 1886 that made him famous, and led to the general acceptance of the periodic table. In all, Mendeleev gave Sanskrit names to eight elements in his periodic table. This note presents the connection between the Sanskrit tradition and the crucial insight that led him to his discovery.

Mendeleev's periodic table of elements, formulated in 1869, is one of the major conceptual advances in the history of science. Dmitri Mendeleev (1834-1907) arranged in the table the 63 known elements based on atomic weight, which he published in his article "On the Relationship of the Properties of the Elements to their Atomic Weights"[1]. He left space for new elements, and predicted three yet-to-be-discovered elements including *eka-silicon* and *eka-boron*. It is the Sanskrit "eka" of these names that we wish to investigate in this note.

Mendeleev was born at Tobolsk, Siberia, and educated in St. Petersburg. He was appointed to a professorship in St. Petersburg 1863 and in 1866 he succeeded to the Chair of Chemistry in the University. He is best known for his work on the periodic table, which was soon recognized since he predicted the existence and properties of new elements and indicated that some accepted atomic weights of the then known elements were in error. His table was an improvement on the classification by Beguyer de Chancourtois and Newlands and was published a year before the work of Lothar Meyer.

The earlier attempts at classification had considered some two-dimensional schemes, but they remained arbitrary in their conception. Mendeleev's main contribution was his insistence that the two-dimensional arrangement was comprehensive. In this he appears to have been inspired by the two-dimensional arrangement of Sanskrit sounds, which he indirectly acknowledges in his naming scheme. But there exist two different arrangements of Sanskrit sounds, and the purpose of this note is to see which one of these was the source of his inspiration.

## 2. Mendeleev's 1869 Paper and later Work

Here's an English translation of his brief paper: "By ordering the elements according to increasing atomic weight in vertical rows so that the horizontal rows contain analogous elements, still ordered by increasing atomic weight, one obtains the following arrangement, from which a few general conclusions may be derived.

**Table 1.** Mendeleev's 1869 periodic table

|   |   |   |   |   |   |
|---|---|---|---|---|---|
|   |   |   | Ti=50 | Zr=90 | ?=180 |
|   |   |   | V=51 | Nb=94 | Ta=182 |
|   |   |   | Cr=52 | Mo=96 | W=186 |
|   |   |   | Mn=55 | Rh=104,4 | Pt=197,4 |
|   |   |   | Fe=56 | Ru=104,4 | Ir=198 |
|   |   |   | Ni=Co=59 | Pd=106,6 | Os=199 |
| H=1 |   |   | Cu=63,4 | Ag=108 | Hg=200 |
|   | Be=9,4 | Mg=24 | Zn=65,2 | Cd=112 |   |
|   | B=11 | Al=27,4 | **?=68** | Ur=116 | Au=197? |
|   | C=12 | Si=28 | **?=70** | Sn=118 |   |
|   | N=14 | P=31 | As=75 | Sb=122 | Bi=210? |
|   | O=16 | S=32 | Se=79,4 | Te=128? |   |
|   | F=19 | Cl=35,5 | Br=80 | J=127 |   |
| Li=7 | Na=23 | K=39 | Rb=85,4 | Cs=133 | Tl=204 |
|   |   | Ca=40 | Sr=87,6 | Ba=137 | Pb=207 |
|   |   | **?=45** | Ce=92 |   |   |
|   |   | ?Er=56 | La=94 |   |   |
|   |   | ?Yt=60 | Di=95 |   |   |
|   |   | ?In=75,6 | Th=118? |   |   |

1. The elements, if arranged according to their atomic weights, exhibit a periodicity of properties.
2. Chemically analogous elements have either similar atomic weights (Pt, Ir, Os), or weights which increase by equal increments (K, Rb, Cs).
3. The arrangement according to atomic weight corresponds to the *valence* of the element and to a certain extent the difference in chemical behavior, for example Li, Be, B, C, N, O, F.
4. The elements distributed most widely in nature have *small* atomic weights, and all such elements are marked by the distinctness of their behavior. They are, therefore, the *representative* elements; and so the lightest element H is rightly chosen as the most representative.

5. The *magnitude* of the atomic weight determines the properties of the element. Therefore, in the study of compounds, not only the quantities and properties of the elements and their reciprocal behavior is to be taken into consideration, but also the *atomic weight* of the elements. Thus the compounds of S and Tl [Te was intended], Cl and J, display not only many analogies, but also striking differences.
6. One can predict the discovery of many *new* elements, for example analogues of Si and Al with atomic weights of 65-75.
7. A few atomic weights will probably require correction; for example Te cannot have the atomic weight 128, but rather 123-126.
8. From the above table, some new analogies between elements are revealed. Thus Bo (?) [apparently Ur was intended] appears as an analogue of Bo and Al, as is well known to have been long established experimentally."

Mendeleev's textbook, *Osnovy Khimii* (*Principles of Chemistry*; first edition, 1871), described his table at greater length. Table 2 shows the evolution of his arrangement. The full list of his predicted elements together with the Sanskrit names he chose is given in Table 3.

**Table 2.** Mendeleev's 1872 version of the periodic table

| REIHEN | GRUPPE I. — $R_2O$ | GRUPPE II. — $RO$ | GRUPPE III. — $R_2O_3$ | GRUPPE IV. $RH^4$ $RO_2$ | GRUPPE V. $RH^3$ $R_2O_5$ | GRUPPE VI. $RH^2$ $RO_3$ | GRUPPE VII. $RH$ $R_2O_7$ | GRUPPE VIII. — $RO_4$ |
|---|---|---|---|---|---|---|---|---|
| 1 | H=1 | | | | | | | |
| 2 | Li=7 | Be=9,4 | B=11 | C=12 | N=14 | O=16 | F=19 | |
| 3 | Na=23 | Mg=24 | Al=27,3 | Si=28 | P=31 | S=32 | Cl=35,5 | |
| 4 | K=39 | Ca=40 | —=44 | Ti=48 | V=51 | Cr=52 | Mn=55 | Fe=56, Co=59, Ni=59, Cu=63. |
| 5 | (Cu=63) | Zn=65 | —=68 | —=72 | As=75 | Se=78 | Br=80 | |
| 6 | Rb=85 | Sr=87 | ?Yt=88 | Zr=90 | Nb=94 | Mo=96 | —=100 | Ru=104, Rh=104, Pd=106, Ag=108. |
| 7 | (Ag=108) | Cd=112 | In=113 | Sn=118 | Sb=122 | Te=125 | J=127 | |
| 8 | Cs=133 | Ba=137 | ?Di=138 | ?Ce=140 | — | — | — | — — — — |
| 9 | (—) | — | — | — | — | — | — | |
| 10 | — | — | ?Er=178 | ?La=180 | Ta=182 | W=184 | — | Os=195, Ir=197, Pt=198, Au=199. |
| 11 | (Au=199) | Hg=200 | Tl=204 | Pb=207 | Bi=208 | — | — | |
| 12 | — | — | — | Th=231 | — | U=240 | — | — — — — |

**Figure 2.5** Dmitri Mendeleev's 1872 periodic table. The spaces marked with blank lines represent elements that Mendeleev deduced existed but were unknown at the time, so he left places for them in the table. The symbols at the top of the columns (e.g., $R^2O$ and $RH^4$) are molecular formulas written in the style of the 19th century.

**Table 3:** The Full List of Mendeleev's Predictions with their Sanskrit Names

*Mendeleev's Given Name*             *Modern Name*

| Eka-aluminium | Gallium |
|---|---|
| Eka-boron | Scandium |
| Eka-silicon | Germanium |
| Eka-manganese | Technetium |
| Tri-manganese | Rhenium |
| Dvi-tellurium | Polonium |
| Dvi-caesium | Francium |
| Eka-tantalum | Protactinium |

Julius Lothar Meyer (1830–1895) published his classic paper of 1870 [2] that also presented the periodicity of atomic volume plotted against atomic weight. Meyer and Mendeleev carried on a long drawn-out dispute over priority. But it was Mendeleev's predictions of yet-unknown elements that secured his fame. The most famous of these predictions was for *eka*-silicon (germanium) for which not only did he postulate its existence, but also its properties in its chloride and oxide combinations. Below is his paper in the *Journal of the Russian Chemical Society,* **3:** 25-56 (1871), where he made predictions regarding eka-boron (scandium).

<div style="text-align:center">

A NATURAL SYSTEM OF THE ELEMENTS AND ITS USE IN PREDICTING
THE PROPERTIES OF UNDISCOVERED ELEMENTS

</div>

"And now, in order to clarify the matter further, I wish to draw some conclusions as to the chemical and physical properties of those elements which have not been placed in the system and which are still undiscovered but whose discovery is very probable. I think that until now we have not had any chance to foresee the absence of these or other elements, because we have had no order for their arrangement, and even less have we had occasion to predict the properties of such elements. An established system is limited by its order of known or discovered elements. With the periodic and atomic relations now shown to exist between all the atoms and the properties of their elements, we see the possibility not only of noting the absence of some of them but even of determining, and with great assurance and certainty, the properties of these as yet unknown elements; it is possible to predict their atomic weight, density in the free state or in the form of oxides, acidity or basicity, degree of oxidation, and ability to be reduced and to form double salts and to describe the properties of the metalloorganic compounds and chlorides of the given element; it is even possible also to describe the properties of some compounds of these unknown elements in still greater detail. Although at the present time it is not possible to say when one of these bodies which I have predicted will be discovered, yet the opportunity exists for finally convincing myself and other chemists of the truth of those hypotheses which lie at the base of the system I have drawn up. Personally, for me these assumptions have become so strong that, as in the case of indium, there is justification for the ideas which are based on the periodic law which lies at the base of all this study.

"Among the ordinary elements, the *lack* of a number of *analogues of boron and aluminum* is very striking, that is, in group III, and it is certain that we lack an element of this group immediately following aluminum; this must be found in the even, or second, series, immediately after potassium and calcium. Since the atomic weights of these latter are near 40, and since then in this row the element of group IV, titanium, Ti = 50, follows, then the atomic weight of the missing element should be nearly 45. Since this element belongs to an even series, it should have more basic properties than the lower elements of group III, boron or aluminum, that is, its oxide, $R_2O_3$, should be a stronger base. An indication of this is that the oxide of titanium, $TiO_2$, with the properties of a very weak acid, also shows many signs of being clearly basic. On the basis of these properties, the oxide of the metal should still be weak, like the weakly basic properties of titanium dioxide; compared to aluminum, this oxide should have a rnore strongly basic character, and therefore, probably, it should not decompose water, and it should combine with acids and alkalis to form simple salts; ammonia will not dissolve it, but perhaps the hydrate will dissolve weakly in potassium hydroxide, although the latter is doubtful because the element belongs to the even series and to a group of elements whose oxides contain a small amount of oxygen. I have decided to give this element the preliminary name of *ekaboron,* deriving the name from this, that it follows boron as the first element of the even group, and the syllable *eka* comes from the Sanskrit word meaning "one." Eb = 45. Ekaboron should be a metal with an atomic volume of about 15, because in the elements of the second series, and in all the even series, the atomic volume falls quickly as we go from the first group to the following ones. Actually, the volume of potassium is nearly 50, calcium nearly 25, titanium and vanadium nearly 9, and chromium, molybdenum, and iron nearly 7; thus the specific gravity of the metal should be close to 3.0, since its atomic weight 45. The metal will be nonvolatile, because all the metals in the even series of all the groups (except group I) are nonvolatile; hence it can hardly be discovered by the ordinary method of spectrum analysis. It should not decompose water at ordinary temperature, but at somewhat raised temperatures it should decompose it, as do many other metals of this series which form basic oxides. Finally, it will dissolve in acids. Its chloride $EbCl_3$ (perhaps $Eb_2Cl_6$), should be a volatile substance but a salt, since it corresponds to a basic oxide. Water will act on it as it does on the chlorides of calcium and magnesium, that is, ekaboron chloride will be a hygroscopic body and will be able to evolve hydrogen chloride without having the character of a hydrochloride. Since the volume ot calcium chloride = 49 and that of titanium chloride = 109, the volume of ekaboron chloride should be close to 78, and therefore its specific gravity will probably be about 2.0 Ekaboron oxide, $Eb_2O_3$, should be a nonvolatile substance and probably should not fuse; it should be insoluble in water, because even calcium oxide is very slightly soluble in water, but it will probably dissolve in acid. Its specific volume should be about 39, because in the series potassium oxide has a volume of 35, CaO = 18, TiO = 20, and $CrO_8$ = 36; that is, considered on the basis of a content of one atom of oxygen, the volume quickly falls to the right, thus, for potassium = 35, for calcium = 18, for titanium = 10, for chromium = 12, and therefore the volume for ekaboron oxide containing one atom of oxygen should be nearly 13, and so the formula $Eb_2O_3$ should correspond to a volume of about 39, and therefore anhydrous ekaboron oxide will have a specific gravity close to 3.5. Since it is a sufficiently strong base, this oxide should show little tendency to form alums, although it will probably give alum-forming compounds, that is, double salts with potassium sulfate. Finally, ekaboron will not form metalloorganic compounds, since it is one of the metals of an even series. Judging by the data now known for the elements which accompany cerium, none of them belong in the place which is assigned to ekaboron, so that this metal is certainly not one of the members of the cerium complex which is now known."

**3. The Sanskrit Tradition and Mendeleev's Discovery**

Note that the Sanskrit alphabet is represented traditionally in a two-dimensional arrangement based on how the sounds are produced (Tables 4).

**Table 4:** The Sanskrit alphabet in the Devanāgarī script

| | |
|---|---|
| अ आ इ ई उ ऊ | a ā i ī u ū |
| ऋ ॠ ऌ ॡ | ṛi ṛī lṛi lṛī |
| ए ऐ ओ औ अं अः | e ai o au am ah |
| | |
| क ख ग घ ङ | k kh g gh ṅ |
| च छ ज झ ञ | c ch j jh ñ |
| ट ठ ड ढ ण | ṭ ṭh ḍ ḍh ṇ |
| त थ द ध न | t th d dh n |
| प फ ब भ म | p ph b bh m |
| य र ल व | y r l v |
| श ष स ह | ś ṣ s h |

The first group of sixteen is that of the vowels, which are simple vowels or diphthongs. The remaining letters are consonants which are divided into five classes: those pronounced from the throat are *gutturals*; those from the palate are *palatals*; those pronounced from the roof of the mouth are *cerebrals*; those pronounced from the teeth are *dentals*; those pronounced from the lips are *labials*. Each of these classes contains seven consonants: five mutes, one semi-vowel, and one sibilant.

Pānini, the author of a famed grammar of Sanskrit who lived in the fifth century BC, in his Śiva Sūtras (also called Māheśvara Sūtras) came up with another classification in 14 categories based on phonological properties of sounds:

**Table 5:** The Śiva Sūtra

1. अ इ उण्

2. ऋ लृक्

3. ए ओङ्

4. ऐ औच्

5. ह य व रट्

6. लण्

7. ञ म ङ ण नम्

8. झ भञ्

9. घ ढ धष्

10. ज ब ग ड दश्

11. ख फ छ ठ थ च ट तव्

12. क पय्

13. श ष सर्

14. हल्

According to Professor Paul Kiparsky of Stanford University, Mendeleev was a friend and colleague of the Sanskritist Böhtlingk, who was preparing the second edition of his book on Panini [3] at about this time, and Mendeleev wished to honor Pānini with his nomenclature.

Noting that there are striking similarities between the Periodic Table and the introductory Śiva Sūtras in Panini's grammar, Kiparsky says [4]:

> [T]he analogies between the two systems are striking. Just as Panini found that the phonological patterning of sounds in the language is a function of their articulatory properties, so Mendeleev found that the chemical properties of elements are a function of their atomic weights. Like Panini, Mendeleev arrived at his discovery through a search for the "grammar" of the elements (using what he called the principle of isomorphism, and looking for general formulas to generate the possible chemical compounds). Just as Panini arranged the sounds in order of increasing phonetic complexity (e.g. with the simple stops k,p... preceding the other stops, and representing all of them in expressions like kU, pU) so Mendeleev arranged the elements in order of increasing atomic weights, and called the first row (oxygen, nitrogen, carbon etc.) "typical (or representative) elements". Just as Panini broke the phonetic parallelism of sounds when the simplicity of the system required it, e.g. putting the velar to the right of the labial in the nasal row, so Mendeleev gave priority to isomorphism over atomic weights when they conflicted, e.g. putting beryllium in the magnesium family because it patterns with it even though by atomic weight it seemed to belong with nitrogen and phosphorus. In both cases, the periodicities they discovered would later be explained by a theory of the internal structure of the elements.

Kiparsky has examined the question of the optimality of the Śiva Sutras elsewhere [5]. He suggests that this optimality might have provided him with the confidence in a similarly optimal two-dimensional table of elements.

My own view is that it is unlikely that Panini's Śiva Sutras that influenced him, because there is no evidence that he knew Sanskrit well enough to appreciate the subtle points related to the organization of the Śiva Sutras. It is more plausible that he noted the comprehensiveness of the two-dimensional arrangement of the Sanskrit alphabet (*varnamālā*) which is apparent to even the beginning student of the language. The tabular form of the Sanskrit letters is due to the two parameters (point of articulation and aspiration) at the basis of the sounds, and Mendeleev must have recognized that ratios/valency and atomic weight likewise defined a two-dimensional basis for the elements.

Convinced that the analogy was fundamental, Mendeleev theorized that the gaps that lay in his table must correspond to undiscovered elements. For his predicted eight elements, he used the prefixes of *eka, dvi*, and *tri* (Sanskrit *one, two, three*) in their naming.

It should be recognized that some of the most brilliant European minds studied Sanskrit in the nineteenth century, and philology and natural science papers were published in the same proceedings of the St. Petersburg Academy of Sciences, as at other academies. The two-dimensional regular representation of Sanskrit sounds must have been well-known to Mendeleev. The paths of Böhtlingk and Mendeleev crossed in many ways: Mendeleev lectured at the Academy when he was awarded its Demidov prize for his book *Organic Chemistry*, which had appeared in 1861, when Böhtlingk was on the nomination committee for the prize.

Mendeleev, by using Sanskrit names, was tipping his hat to the Sanskrit grammarians of yore, who had created astonishingly sophisticated theories of language based on their discovery of the two-dimensional patterns in basic sounds.

## 4. The Indian Atomic Tradition

It should be mentioned that Indian physical ideas considered all matter to be created of the same indestructible atoms, which combines in diverse ways to generate different substances. The objective elements of the physical world are characterized by substance, quality, and action. There is a further characterization in terms of non-reactive and reactive properties [6].

In one of the Indian conceptions, two atoms combine to form a binary molecule, and two, three, four or more binary molecules combine into grosser molecules of different substances. The other view is that atoms form dyads and triads directly to form molecules of different substances. Atoms possess an incessant vibratory motion.

Molecules can also break up under the influence of heat. Heat and light rays are taken to consist of very small particles of high velocity. Being particles, their velocity is finite. The particles of heat and light can be endowed with different characteristics and so heat and light can be of different kinds.

But these ideas were mainly debated in philosophical commentaries, and they were not developed further to find their inner contradictions, so as to lead to a satisfactory picture of the diversity of substances in nature. Had that been done, a much more comprehensive picture of matter would have emerged in ancient India.

## 5. Concluding Remarks:

Mendeleev anticipated Andrews' concept (1869) of the critical temperature of gases. He also investigated the thermal expansion of liquids, and studied the nature and origin of petroleum. He was considered one of the great teachers of his time. Resigning his professorship in 1890, three years later he became director of the bureau of weights and measures in St. Petersburg, where he remained until his death.

In 1913, almost fifty years after Mendeleev, Henry Moseley (1887-1915) measured the wavelengths of the X-ray spectral lines of many elements, showing that the ordering of the wavelengths of the X-ray emissions coincided with the ordering of the elements by atomic number. With the discovery of isotopes of the elements, it became apparent that atomic mass was not the significant player in the periodic law as Mendeleev, Meyers and others had proposed, but rather, the properties of the elements varied periodically with atomic number. But since this was a minor modification, the table has continued to be called after Mendeleev.

The connections between computer science and Sanskrit grammatical conception have been investigated by several scholars [7]. But the connection between these grammatical ideas and modern theories of matter is a most fascinating chapter of history of science that has remained forgotten for over a hundred and thirty years.